\documentclass[osajnl2,amsmath,amssymb, aps, pra,twocolumn,showpacs,superscriptaddress]{revtex4-1}%CM

\usepackage{graphicx}% Include figure files
\usepackage{dcolumn}% Align table columns on decimal point
\usepackage{bm}% bold math
\usepackage{epstopdf}
\usepackage{color}
\begin{document}

%\preprint{APS/123-QED}

\title{Underwater acoustic wave generation by filamentation of terawatt ultrashort laser pulses}% Force line breaks with \\

%\author{Vytautas Jukna}
% \email{vytautas.jukna@ensta-paristech.fr}
%\author{Amelie Jarnac}
%\author{Yohann Brelet}
%\author{J\'{e}r\^{o}me Carbonnel}
%\author{Yves-Bernard Andr\'{e}}
%\author{Andr\'{e} Mysyrowicz}
%\author{Aur\'{e}lien Houard}
% \affiliation{LOA, ENSTA-ParisTech, CNRS, Ecole Polytechnique, Universit\'{e} Paris Saclay, 91762 Palaiseau cedex, France}
%\author{Carles Mili\'{a}n}
%\author{Arnaud Couairon}
% \affiliation{Centre de Physique Th\'{e}orique, CNRS, Ecole polytechnique, Universit\'{e} Paris Saclay, F-91128 Palaiseau, France}%
%\author{R\'{e}gine Guillermin}
%\author{Jean-Pierre Sessarego}
% \affiliation{Laboratoire de M\'{e}canique et d’Acoustique, 31 Chemin Joseph Aiguier, 13402 Marseille cedex 20, France}%%Lines break automatically or can be forced with
%\author{Dominique Fattaccioli}
% \affiliation{DGA Techniques Navales, Toulon, France}%Lines break automatically or can be forced with

\author{Vytautas Jukna}
\email{vytautas.jukna@ensta-paristech.fr}
 \affiliation{LOA, ENSTA-ParisTech, CNRS, Ecole Polytechnique, Universit\'{e} Paris Saclay, 91762 Palaiseau cedex, France}
\author{Amelie Jarnac}
\affiliation{LOA, ENSTA-ParisTech, CNRS, Ecole Polytechnique, Universit\'{e} Paris Saclay, 91762 Palaiseau cedex, France}
\author{Carles Mili\'{a}n}
\affiliation{Centre de Physique Th\'{e}orique, CNRS, Ecole polytechnique, Universit\'{e} Paris Saclay, F-91128 Palaiseau, France}
\author{Yohann Brelet}
\affiliation{LOA, ENSTA-ParisTech, CNRS, Ecole Polytechnique, Universit\'{e} Paris Saclay, 91762 Palaiseau cedex, France}
\author{J\'{e}r\^{o}me Carbonnel}
\affiliation{LOA, ENSTA-ParisTech, CNRS, Ecole Polytechnique, Universit\'{e} Paris Saclay, 91762 Palaiseau cedex, France}
\author{Yves-Bernard Andr\'{e}}
\affiliation{LOA, ENSTA-ParisTech, CNRS, Ecole Polytechnique, Universit\'{e} Paris Saclay, 91762 Palaiseau cedex, France}
\author{R\'{e}gine Guillermin}
\affiliation{Laboratoire de M\'ecanique et d’Acoustique, CNRS - UPR 7051, 4 impasse Nikola Tesla, CS 40006, 13453 Marseille Cedex 13}
\author{Jean-Pierre Sessarego}
\affiliation{Laboratoire de M\'ecanique et d’Acoustique, CNRS - UPR 7051, 4 impasse Nikola Tesla, CS 40006, 13453 Marseille Cedex 13}
\author{Dominique Fattaccioli}
\affiliation{DGA Techniques Navales, Toulon, France}%Lines break automatically or can be forced with
\author{Andr\'{e} Mysyrowicz}
\affiliation{LOA, ENSTA-ParisTech, CNRS, Ecole Polytechnique, Universit\'{e} Paris Saclay, 91762 Palaiseau cedex, France}
\author{Arnaud Couairon}
\affiliation{Centre de Physique Th\'{e}orique, CNRS, Ecole polytechnique, Universit\'{e} Paris Saclay, F-91128 Palaiseau, France}%
%Lines break automatically or can be forced with
\author{Aur\'{e}lien Houard}
\affiliation{LOA, ENSTA-ParisTech, CNRS, Ecole Polytechnique, Universit\'{e} Paris Saclay, 91762 Palaiseau cedex, France}

%\author{C. Mili\'{a}n}
%\email{carles.milian@cpht.polytechnique.fr}
%\affiliation{Centre de Physique Th\'{e}orique, CNRS, \'{E}cole Polytechnique, F-91128 Palaiseau, France}
%\author{A. V. Gorbach}
%\affiliation{Department of Physics, University of Bath, Bath BA2 7AY, United Kingdom}
%\author{M. Taki}
%\affiliation{PhLAM, Universit\'{e} de Lille 1, F-59655 Villeneuve d'Ascq Cedex, France}
%\author{A. V. Yulin}
%\affiliation{ITMO University, Kronverksky pr. 49, St. Petersburg,  197101, Russian Federation}
%\author{D. V. Skryabin}
%\affiliation{Department of Physics, University of Bath, Bath BA2 7AY, United Kingdom}
%\affiliation{ITMO University, Kronverksky pr. 49, St. Petersburg,  197101, Russian Federation}

\date{\today}% It is always \today, today,

\begin{abstract}
Acoustic signals generated by filamentation of ultrashort TW laser pulses in water are characterized experimentally. Measurements reveal a strong influence of input pulse duration on the shape and intensity of the acoustic wave. Numerical simulations of the laser pulse nonlinear propagation and the subsequent water hydrodynamics and acoustic wave generation show that the strong acoustic emission is related to the mechanism of superfilamention in water. The elongated shape of the plasma volume where energy is deposited drives the far-field profile of the acoustic signal, which takes the form of a radially directed pressure wave with a single oscillation and a very broad spectrum.
%\begin{description}
%%\item[PACS numbers]
%%%May be entered using the \verb+\pacs{#1}+ command.
%%42.65.Jx Beam trapping, self-focusing and defocusing; self-phase modulation
%%43.30.+m Underwater sound
%%47.40.-x Compressible flows; shock waves
%\end{description}
\end{abstract}

\maketitle

%\tableofcontents

\section{\label{sec:Introduction}Introduction}
%\textbf{There is a concern that introduction uses cavitation and bubbles while we do not say anything about them in the actual paper. I do not see the problem as we have other paragraphs which explains what we did in this paper. Carles say that it might be a problem with hydrodynamical physics community. What do you think?}
%\textcolor[rgb]{1.00,0.00,0.00}{
When a compressible liquid submitted to external forces ruptures violently, cavitation occurs and nucleates bubbles that subsequently implode and undergo oscillations driven by the external fluid pressure in the surrounding liquid. An acoustic signal is released from the bubble implosion. Cavitation and acoustic wave generation can be a phenomenon to avoid or in contrast, a desired effect provided a certain degree of control can be reached. For instance, cavitation is well known to induce damage on ship propellers, but cavitation-induced high-velocity jets and high pressure acoustic wave in water allow snapping shrimps to stun or kill prey animals \cite{Versluis2000}. Not only in the natural world but also for numerous applications, from chemical engineering, biomedical ultrasound imaging, to mechanical optical cleaning \cite{Lauterborn2013}, internal combustion engine efficiency, and interface science \cite{Lauterborn2010}, would it be desirable to control cavitation and subsequently pressure wave release.

Laser-induced energy deposition in water and effects following optical breakdown have been investigated for the past decades (see \cite{Linz2015} for recent findings). Laser induced cavitation in water was discovered  in the early sixties \cite{Askaryan1963,White1963} and has been the subject of continued interest as it was rapidly recognized that the development of laser-induced acoustic sources in water could open up new possibilities for underwater communications, for high resolution imaging, tomography and fast characterization of marine environment with the aim of exploiting sea resources,  or for remote acoustic control of submerged instruments \cite{Egerev2003}. The first experiments were performed with long pulse laser sources, leading to a slow heating of water followed by thermal expansion and emission of an acoustic wave \cite{Chotiros1988,Tang1991,Blackmon}. The conversion efficiencies from light to the acoustic signal was however reported to be enhanced with nanosecond laser pulses, leading to optical breakdown, rapid heating of the focal volume producing pressures in the gigapascal range and explosive expansion followed by the emission of a shock wave \cite{Noack1999,Vogel1999,Potemkin2014}. Femtosecond laser pulses open up new possibilities in this field as they were recently shown to lead to ultrabroad acoustic signals \cite{Brelet2015}. The nonlinear propagation of femtosecond laser pulses in gases or liquids leads to light-plasma filaments, where the laser beam shrinks upon itself due to the Kerr nonlinearity, to reach intensity levels that exceed the threshold for optical field ionization \cite{Couairon2007}. This high intensity can be sustained over extended distances and the filament itself can be generated remotely, adding to the potential flexibility in tuning laser-induced acoustic sources. The dynamics of femtosecond filamentation in water and its various properties has been investigated thoroughly in the past decade \cite{Dubietis2006,Minardi2008,Minardi2009,Sreeja2013,Jarnac2014,Minardi2014,Milian2014}, however only a few investigations focus on the potential of filaments for cavitation or acoustic wave generation \cite{Chin1996,FaccioPRE2012,FaccioPRA2012,Jones2006,Jones2007,Potemkin2014,Milian2015b}. In particular Potemkin et al. demonstrated enhancement of the acoustic signal amplitude with an increase in the length of the focal region \cite{Potemkin2015}.

In this paper, we present investigations on acoustic signals generated by ultrashort laser pulse filamentation. Acoustic measurements were done utilizing femtosecond and picosecond laser pulses with multi milli Joule energy as a source of acoustic signals. Numerical simulations are performed for understanding the nonlinear propagation of the laser beam through water, the subsequent hydrodynamic expansion of the focal volume and the propagation of the generated acoustic signal. The numerical simulations are divided into three stages discriminated by the duration of the process: (i) nonlinear propagation of the beam and laser pulse energy deposition into water, (ii) laser-induced nonlinear hydrodynamics and shock-wave formation, (iii) propagation of the acoustic wave to the hydrophone. Acoustic signals recorded at distance from the filament exhibit signatures of the focal volume shape. Our numerical simulations show that a nonsymmetrical acoustic signal arising in conditions for superfilamentation can be interpreted as a manifestation of the shape of the focal volume, which depends on the laser pulse energy and focusing geometry. Loose focusing leads to cylindrical focal regions whereas an increase of the numerical aperture leads to a conically-shaped and shorter focal volume. Ability to dynamically control the directivity of the acoustic sources is important for underwater detection and communications.
\section{\label{sec:Experiment}Experimental setup and results}
\begin{figure}[h]
\includegraphics[width=0.75\columnwidth]{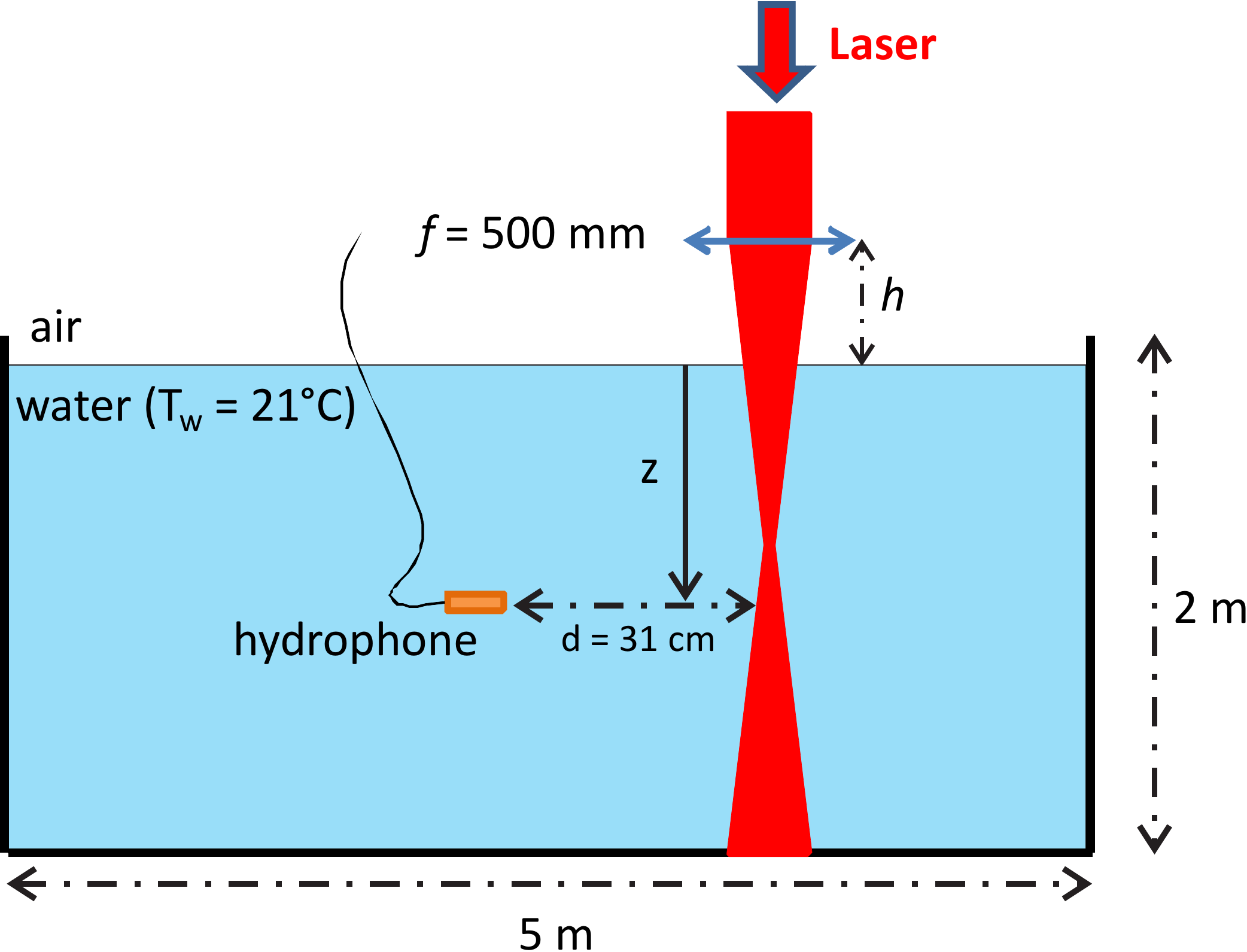}
\caption{\label{fig:Experiment} Experimental setup for generation of acoustic waves with a laser beam and recording them with a hydrophone. $h$ is the distance between the surface of water and the lens while $z$ defines the direction of hydrophone displacement for spatio-temporal acoustic wave analysis.}
\end{figure}

\begin{figure}[h]
\includegraphics[width=0.99\columnwidth]{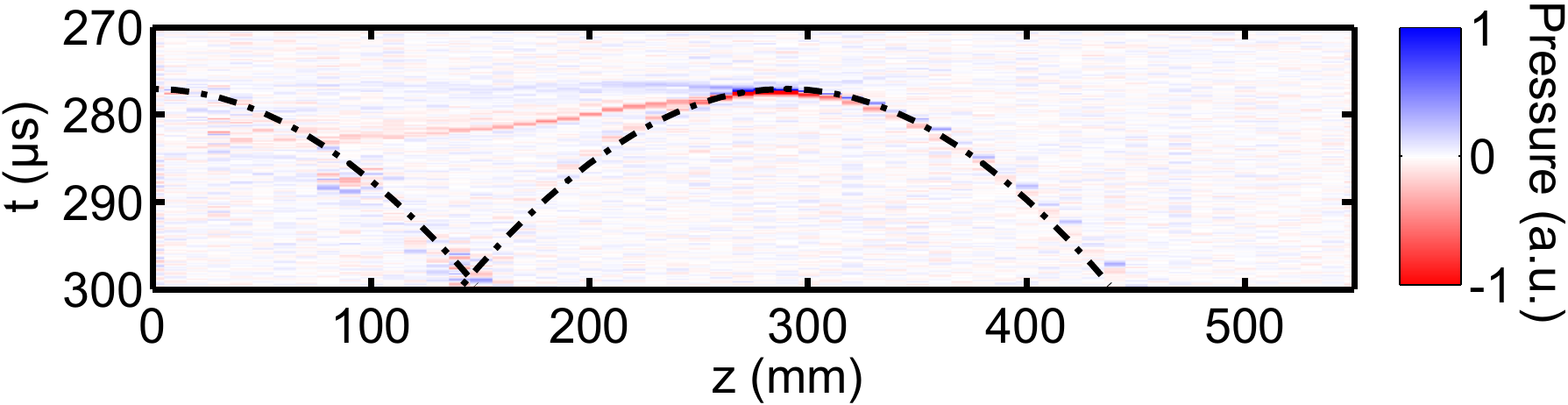}
\caption{\label{fig:Exp1} Typical acoustic wave profile registered by a hydrophone. The focusing lens was 30 cm above the surface of water and the pulse duration was 5 ps.  Each dashed curve (plotted as an eye-guide) represents the profile for a spherical acoustic wave emitted from a point source at the depth corresponding to the top the hyperbolic branch (the depths of 0 and 297 mm correspond to waves emitted from the water surface and focal region, respectively), and propagating at the sound velocity in water, $c_s$=1487 m/s.}
\end{figure}

\begin{figure}[h]
\includegraphics[width=0.99\columnwidth]{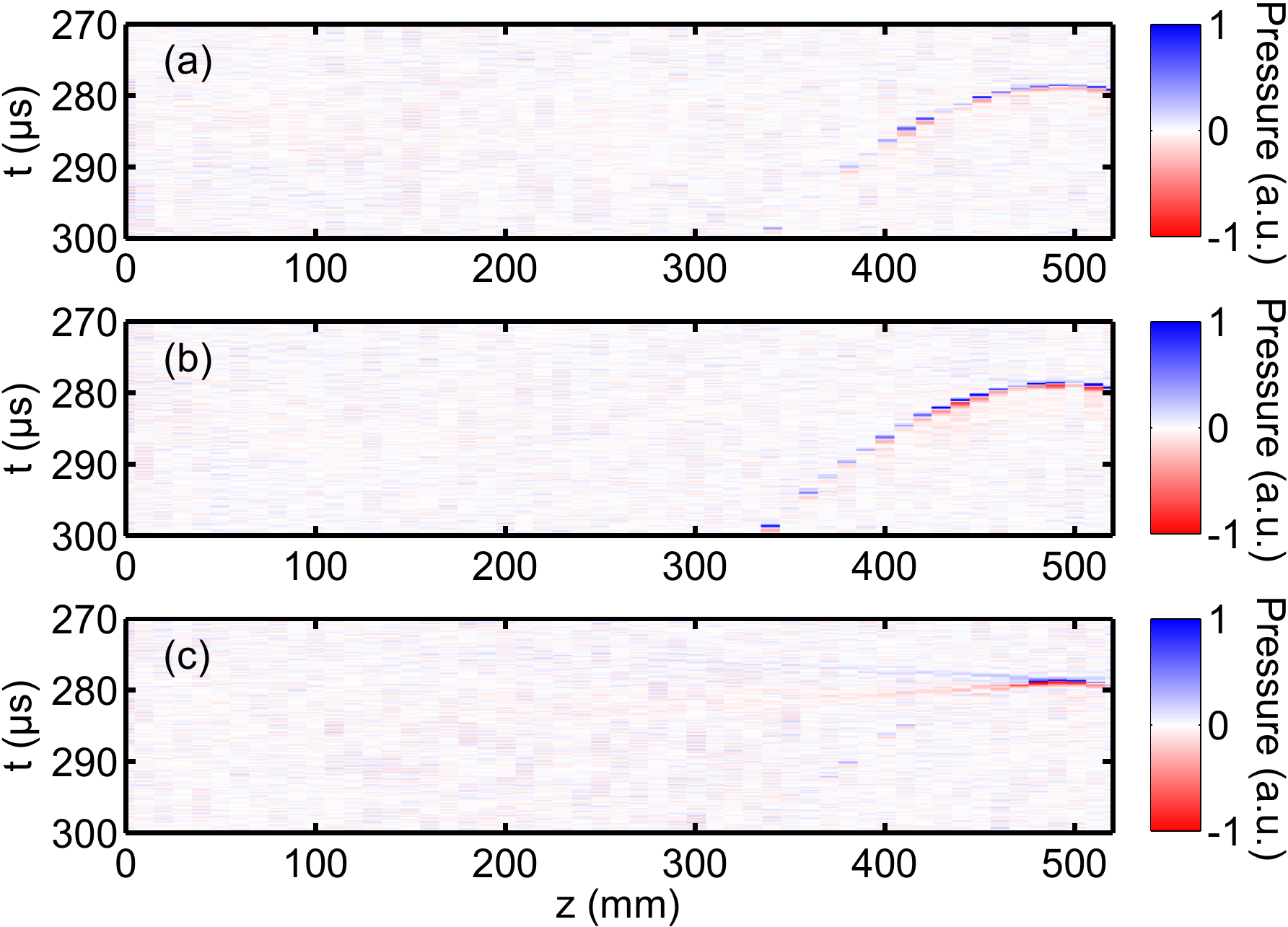}
\caption{\label{fig:Exp2} Comparison of amplitude profiles for the acoustic waves generated by 290 mJ laser pulses with initial duration of (a) 0.25 ps, (b) 0.5 ps, and (c) 5 ps. The lens was positioned at $h$ = 13 cm above the surface of water corresponding to a focus depth of $z_0 \sim$ 500 mm. Acoustic waves are registered by moving the hydrophone along a vertical axis at distance $d$ = 31 cm from the laser propagation axis.}
\end{figure}

The experiment was performed by using a Ti:Sapphire laser with central wavelength of 800 nm, Fourier limited pulse duration of 50 fs and pulse energy of 290 mJ. The beam was focused with a 50 cm lens into a large water tank (Fig. \ref{fig:Experiment}). The lens was placed at a height $h$ above the surface of water. The initial beam diameter (FWHM) on the lens was 35 mm. The pulse duration was changed from 0.25 ps to 5 ps by imposing a linear positive chirp on the 50 fs pulse. To register the acoustic waves emitted by expansion of the focal volume, a very broad band needle hydrophone (flat at $\pm$ 4 dB on the band 200 kHz - 15 MHz) was inserted into water at the separation distance $d$ = 31 cm away from the propagation axis of the laser beam. The spatial-temporal profiles of the acoustic waves were mapped by varying the immersion depth $z$ of the hydrophone, keeping constant the separation distance $d$ and the focusing geometry, and by recording for each depth the acoustic signal reaching the hydrophone after a laser shot.
Figure \ref{fig:Exp1} shows typical measurements. A spherical acoustic wave emitted from a point source located at $z_0$ is expected to reach the hydrophone at a depth $z$ after a delay $t=\sqrt{d^2+(z-z_0 )^2 }/c_s$, where $c_s$ = 1487 m/s denotes the speed of sound in water under normal conditions. In other words, the mapped profile when the focus of the lens is located at $z_0$ should be a hyperbolic branch, centered at $z_0$, as shown by the dashed curves in Fig. \ref{fig:Exp1}. However, our measurements show additional features. A conical (V-shaped) profile is clearly visible as two branches, representing the positive and negative peaks in the acoustic signal intersecting at the emission point (top of the most visible hyperbolic branch at $z_0$ = 270 mm) corresponding to the focus of the lens, where a maximum in signal amplitude is observed. As will be shown below, several regions of the focal volume contribute to the acoustic signal: In addition to the quasi point source at the focus where the plasma density reaches $\sim$ 10$^{22}$ m$^{-3}$, multiple filamentation occurs in the vicinity of the focus in an extended focal volume, featured by focusing conditions, and is responsible for the V-shaped acoustic branches. In this particular case there is another source of acoustic wave located at the surface of water. In the measurements discussed below, we moved the focusing lens closer (separation of $h$ = 13 cm) to the surface of water to prevent interference of this acoustic wave generated at the surface of water with acoustic waves generated in the bulk.
We analyzed acoustic wave generation by filamentation with pulses of different initial pulse durations. Figure \ref{fig:Exp2} shows a comparison of acoustic wave profiles generated with pulse durations from 0.25 to 5 ps. The acoustic waves are plotted with the same colormap for possible relative amplitude comparability. Our measurements revealed that higher amplitudes of acoustic waves tend to be obtained with longer pulse durations when the laser energy is kept constant. In addition the profiles corresponding to the shortest pulse durations (0.25 ps, 0.5 ps) exhibit a single branch as in the case of a point source emitting a spherical wave.  The amplitude profile of the acoustic wave obtained with the longer pulse (5 ps), exhibits the additional V-shaped branches with amplitudes even larger than generated with shorter pulse durations.
\section{\label{sec:Numerics}Numerical simulations of laser energy deposition}
Three different tools were used for numerical simulations of the nonlinear pulse propagation, nonlinear hydrodynamics and generation of the acoustic wave, and its propagation to the hydrophone. Nonlinear propagation of the laser pulse was simulated by means of the code developed for investigating superfilamentation, beam symmetrization in air, and filamentation of large beams from orbit \cite{Point2014,Milian2015,DicaireSubmitted}, which resolves a unidirectional envelope propagation equation describing diffraction, the optical Kerr effect, plasma induced effects including plasma defocusing and nonlinear losses due to its generation by multiphoton and by avalanche ionization (see Appendix A).
%Similar tools were used by various groups \cite{Mlejnek1999, Mechain2004, Berge2004} for multifilament generation analysis.
Our numerical scheme (see \cite{Couairon2011} for details) was upgraded to accommodate beams with high numerical aperture propagating through nonlinear media. A coordinate transformation proposed by Sziklas and Siegman \cite{SziklasAO1975} was implemented, allowing us to easily treat the fast oscillating spatial phase. Our model assumes a fixed Gaussian pulse profile over the whole propagation length. This assumption, associated with a preliminary mapping between peak intensity and electron density through the ionization model allows us to perform (2+1)D simulations with the highly demanding resolution required by our focusing geometry and relatively high pulse energy. With these new features, the code was used to simulate filamentation in water and checked to fairly reproduce experimental findings which will be discussed later in the text. The assumption of a fixed Gaussian pulse shape slightly overestimates energy losses, however the model provides a glimpse in the physics behind laser energy deposition for different input beam conditions.

For the numerically simulated experiments, the lens was located 13 cm above the surface of water. Noise was added to the input beam so as to mimic irregularities on the beam profile and start from realistic initial conditions (as close as possible to experimental conditions). Most of the numerical simulations results in this section deal with a comparison of the laser energy deposition in the focal volume when the duration of the input pulses varies from 0.5 fs to 5 ps, while the pulse energy of 290 mJ is kept constant.

\begin{figure}[h]
\includegraphics[width=0.99\columnwidth]{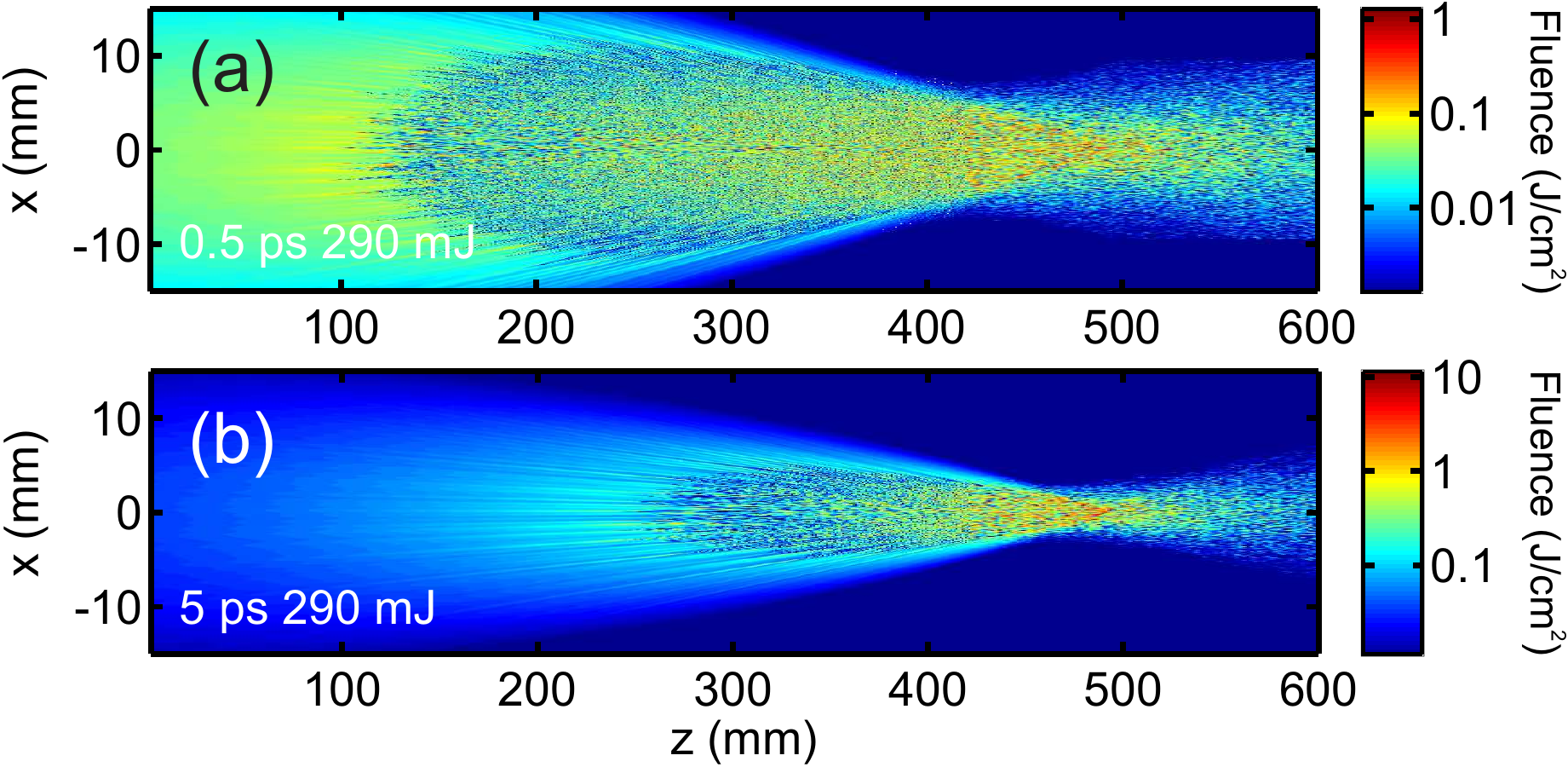}
\caption{\label{fig:Num1} Fluence profiles (cross section along a single transverse dimension $x$) for nonlinear propagation of laser pulses in water. The focus of the lens is at $z$ = 488 cm. Input pulse energy 290 mJ. Pulse duration: (a) 0.5 ps; (b) 5ps.}
\end{figure}

\begin{figure}[!ht]
\includegraphics[width=0.99\columnwidth]{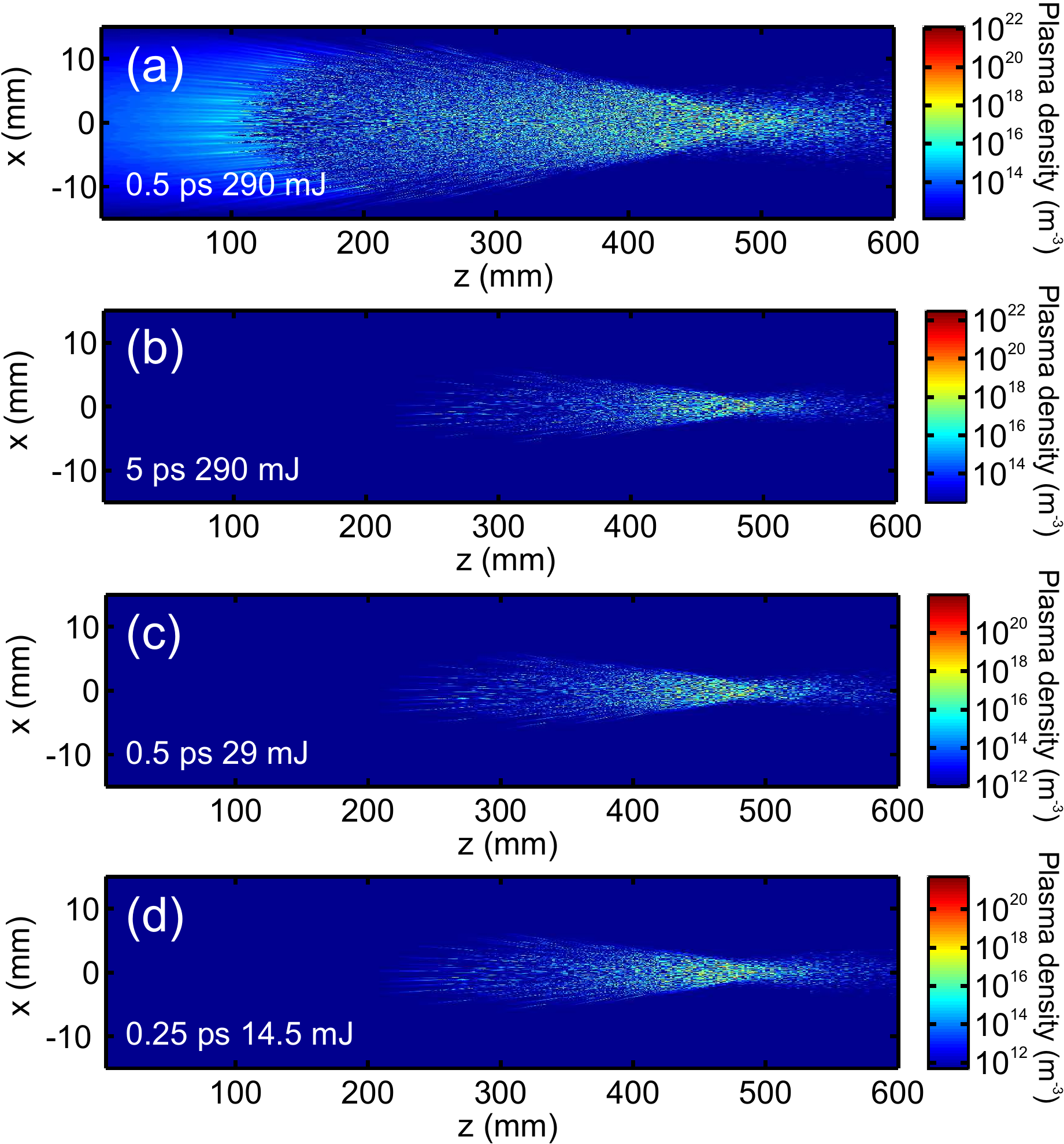}
\caption{\label{fig:Num2} Plasma density profiles in the same conditions as in Fig. \ref{fig:Num1} is depicted in (a) and (b) i.e. when input pulse energy 290 mJ and pulse duration: (a) 0.5 ps; (b) 5 ps. While (b), (c) and (d) are for the cases when input beam peak power was the same for pulse durations 5, 0.5 and 0.25 ps respectively.}
\end{figure}

\begin{figure}[h]
\includegraphics[width=\columnwidth]{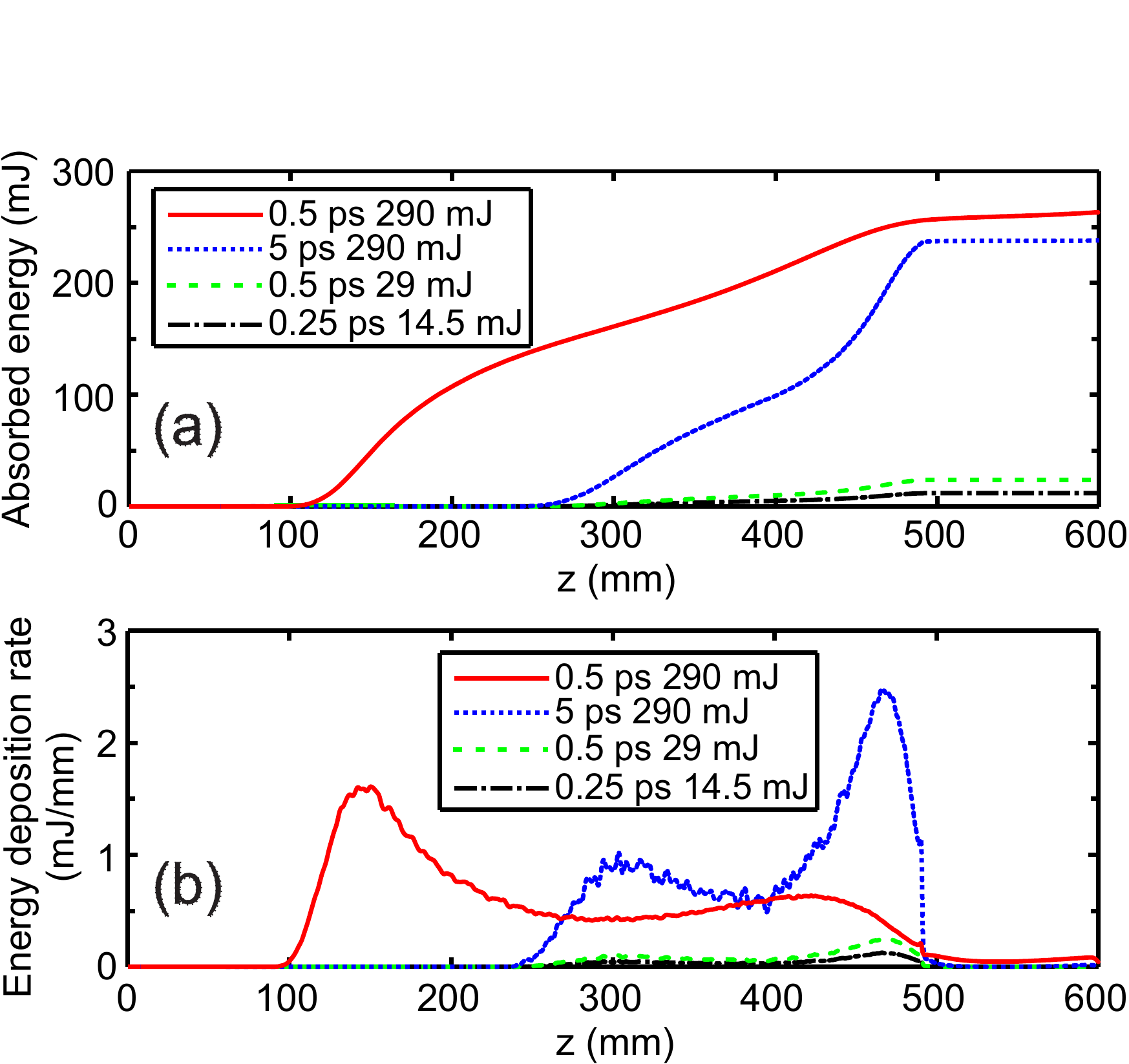}
\caption{\label{fig:NumLines1} Absorbed energy (a) and rate of nonlinear energy losses (b) as a function of propagation distance. The curves correspond to different input pulse energies and pulse durations: 290 mJ, 5 ps (blue); 290 mJ, 0.5 ps (red), 29 mJ, 0.5 ps (green) and 14.5 mJ, 0.25 ps (black).  The blue and red curves correspond to pulses with the same energy while the blue, green and black curves correspond to beams with the same peak power.}
\end{figure}

Figure \ref{fig:Num1} represents a comparison of fluence profiles obtained from numerical simulations with initial pulse durations of 0.5 ps and 5 ps at the same pulse energy 290 mJ. Converging multiple filaments are formed in both cases. The shorter initial pulse initiates filament formation earlier in propagation because the initial peak intensity is 10 times larger and filament generation is directly linked to intensity via modulational instability, which has a maximum growth rate proportional to the intensity. The corresponding plasma density profiles are presented in figure \ref{fig:Num2}(a,b). The plasma volume is larger for the shorter pulse, however, a closer inspection reveals that the plasma is more localized for the 5 ps pulse, and density reaches slightly higher values.
%\textbf{This paragraph was written so that JOSAB could be used as a citation with a slippery proof that longer pulses generates higher plasma density. It is also explaining that pulses with ps duration has higher deposited energy density and gives few additional citations.}
%\textcolor[rgb]{1.00,0.00,0.00}{
This result foresees that the heating of water with ps pulses will be more severe. The existence of optimal pulse (of a few ps's) width that maximizes deposited energy density appears due to local optimization of plasma generation processes (multi photon and avalanche ionization) and beam propagation properties (focusing conditions, self-focusing, plasma defocusing, material dispersion) and is systematically observed in experiments and numerical simulations in dielectrics (see. e.g.,  \cite{Onda2005,Milian2014,Bhuyan2014}).
Figures \ref{fig:Num1} and \ref{fig:Num2}(a,b) compare nonlinear propagation of pulses with the same energy, as in the experiments, resulting in differences in the focal volume mainly due to the different initial peak intensity (power). Figure \ref{fig:Num2}(b,c,d) compares plasma density profiles when the initial peak intensity (peak power) is the same for different initial pulse durations 5 ps, 0.5 ps and 0.25 ps, corresponding to pulse energies of 290, 29 and 14.5 mJ, respectively. In this case the dynamics of multiple filamentation and the features of the plasma volume do not differ significantly. Filaments are generated roughly within the same volume, however, the longest pulse generates plasma at higher density due to a more significant contribution of avalanche ionization.
%\begin{figure}[h]
%\includegraphics[width=\columnwidth]{Num3.jpg}
%\caption{\label{fig:Num3} Plasma density profiles for nonlinear propagation of laser pulses in water. Pulse duration: (a) 0.25 ps; (b) 0.5 ps and (c) 5 ps. The input beam peak power was the same for %all cases.}
%\end{figure}
In order to investigate numerically the propagation of acoustic waves from the focal region to the hydrophone, we analyzed the efficiency of laser energy deposition as a function of pulse duration. Figure \ref{fig:NumLines1}a shows energy losses for all cases discussed previously without separating multiphoton absorption and plasma absorption as both are contributing to locally heat water. Energy transfer to matter is the most important quantity for evaluating heat increase of the matter. The 0.5 ps and 5 ps pulses deposit 89 and 82 \% of their initial energy (290 mJ), respectively. However 0.5 ps pulse starts to lose energy via ionization of water much earlier than the 5 ps pulse. By comparing plasma density plots in figure \ref{fig:Num2}(a,b) it is evident that the plasma volume is also larger for the 0.5 ps pulse. This suggests that the deposited energy density might be lower for the short pulse. Figure \ref{fig:NumLines1}a also shows that by shortening the pulse duration while keeping the peak power fixed, energy loss is decreasing while the plasma generation roughly starts at the same position.
%\begin{figure}[h]
%\includegraphics[width=\columnwidth]{AbsorbedEnergy.jpg}
%\caption{\label{fig:AbsorbedEnergy} Absorbed energy as a function of propagation distance. The curves correspond to different input pulse energies and pulse durations: 290 mJ, 5 ps (blue); 290 mJ, 0.5 ps (red), 29 mJ, 0.5 ps (green) and 14.5 mJ, 0.25 ps (black).  The blue and red curves correspond to pulses with the same energy while the blue, green and black curves correspond to beams with the same peak power.}
%\end{figure}
%\begin{figure}[h]
%\includegraphics[width=\columnwidth]{EnergyDepositionRate.jpg}
%\caption{\label{fig:EnergyDepositionRate} Rate of nonlinear energy losses as a function of propagation distance. Different pulse energies and durations are represented with different colors and are the same as in Fig \ref{fig:AbsorbedEnergy}.}
%\end{figure}

\begin{figure}[h]
\includegraphics[width=\columnwidth]{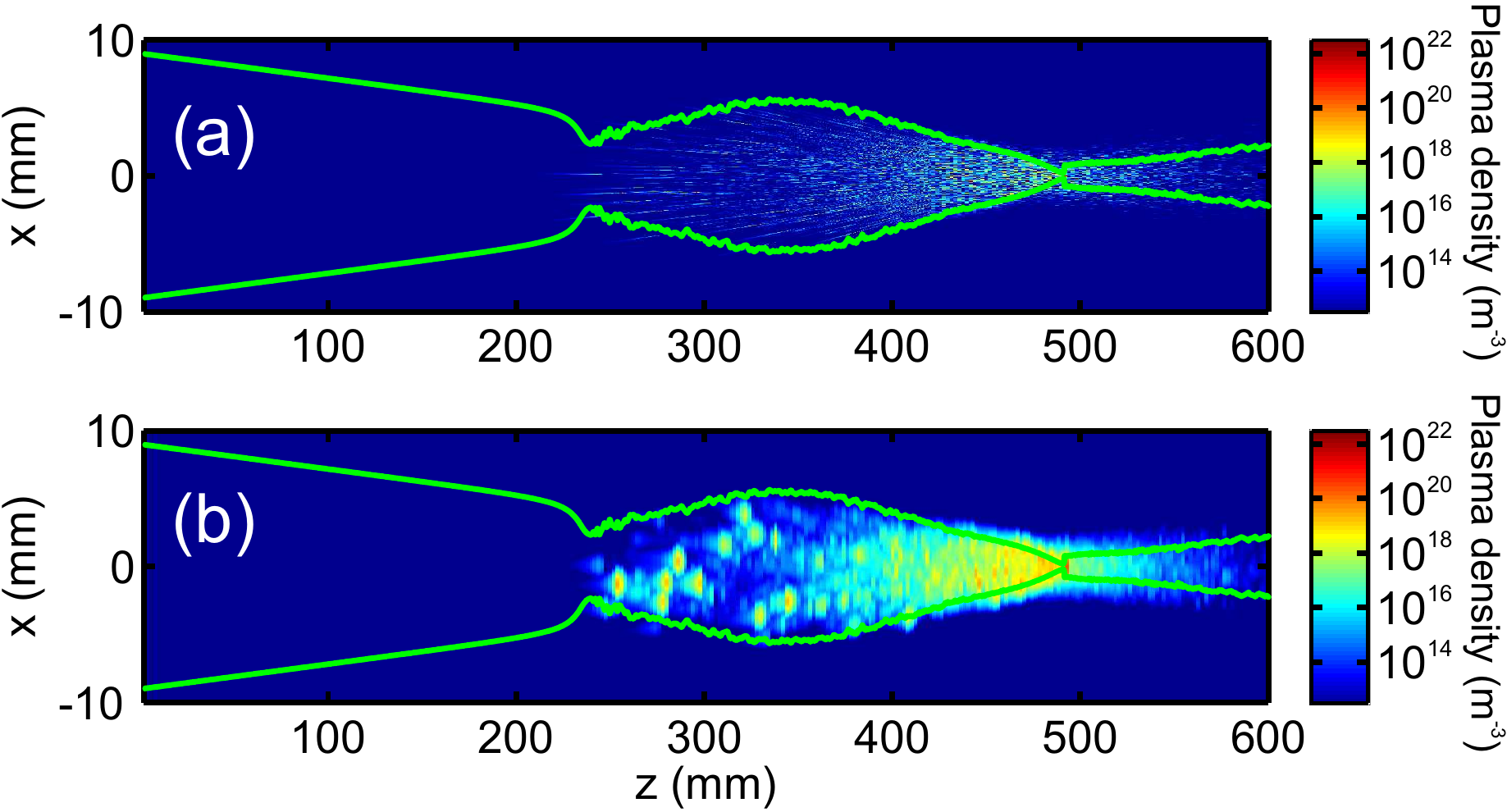}
\caption{\label{fig:Num4} Plasma density distribution from the simulation of pulse propagation with initial energy of 256 mJ and duration of 5 ps (a). The green curve represents the boundary of the focal (plasma) volume. (b) Spectrally filtered plasma density distribution for the same conditions. A log scale is used for both figures.}
\end{figure}

In order to have a diagnostic of the local rate of energy losses, we calculated the derivative of the absorbed energy $d\langle U \rangle/dz$  which represents the energy deposition rate per unit length along the propagation axis
\begin{equation}
\frac{d\langle U \rangle}{dz}=\iiint\limits_{-\infty}^{+\infty}u(x,y,z,t)dxdydt,
\label{eq:EnDepRate}
\end{equation}
where the density of nonlinear losses reads
\begin{eqnarray}
u(x,y,z,t)=&&\sigma\rho_e(x,y,z,t)I(x,y,z,t)+\beta_K I^K\nonumber\\
&&\times(1-\rho_e(x,y,z,t)/\rho_{nt}).
\label{eq:Losses}
\end{eqnarray}
Here $\sigma$=4.7$\times$10$^{-22}$ m$^2$ is the cross section for inverse Bremsstrahlung, $\beta_K$=8.3$\times$10$^{-52}$ cm$^7$/W$^4$ - the multiphoton absorption (MPA) coefficient, $K$=5 - MPA order, $\rho_{nt}$=6.7$\times$10$^{22}$ cm$^{-3}$ - the neutral atom density and $\rho_e(x,y,z,t)$ - the plasma density. The quantity $d\langle U \rangle/dz$  is depicted in figure \ref{fig:NumLines1}(b). For the 0.5 ps pulse, the energy deposition rate exhibits a maximum around $z$ = 15 cm, at the position where multiple filaments form. However, closer to the focus, the energy deposition rate for the 5 ps pulse is the highest. This result already indicates that long pulses deposit energy closer to the linear focus, while the short pulses generates multiple filaments and looses substantial amount of energy long before they reach linear focus. To evaluate the average energy deposited within the focal volume, we evaluate the deposited energy volume by calculating the second order moment $I_2$  of deposited energy assuming a super Gaussian shape:
\begin{widetext}
\begin{equation}
\begin{cases}
I_1=\iiint\limits_{-\infty}^{+\infty}u(x,y,z,t)dzdydt=U_m(z)\int\limits_{0}^{+\infty}\exp\left(-\frac{r^{2s}}{R^{2s}(z)}\right)2\pi r dr=U_m(z)\pi R^2(z)\Gamma\left(1+\frac{1}{s}\right)\\
I_2=\iiint\limits_{-\infty}^{+\infty}\left(x^2+y^2\right) u(x,y,z,t)dzdydt=U_m(z)\int\limits_{0}^{+\infty}r^2\exp\left(-\frac{r^{2s}}{R^{2s}(z)}\right)2\pi r dr=U_m(z)\frac{\pi}{2}R^4(z)\Gamma\left(1+\frac{2}{s}\right)\\
\end{cases}
\label{eq:Moment}
\end{equation}
\end{widetext}
where $\Gamma$ denotes the Gamma functions.
From the numerical evaluation of the deposited energy rate $I_1$ and the second order moment $I_2$, the radius $R(z)$ of the energy deposition volume and the amplitude $U_m$ of the averaged absorbed energy density can be calculated :
\begin{equation}
\begin{cases}
R^2(z)=\frac{2I_2(z)\Gamma(1+1/s)}{I_1(z)\Gamma(1+2/s)}\\
U_m(z)=\frac{I_1^2(z)\Gamma(1+2/s)}{2\pi I_2(z)\Gamma(1+1/s)^2}.
\end{cases}
\label{eq:RUm}
\end{equation}
The radius of the energy deposition volume calculated from the distribution of deposited energy is depicted in figure \ref{fig:Num4} by green curves. The plasma is not homogeneous in this region, reflecting the hot spots generated by multiple filamentation. A spectral filtering technique was used to characterize energy deposition at the meso-scale level, intermediate between the micro plasma channels and the entire focal volume. Figure  \ref{fig:Num4}b shows the locally averaged plasma density obtained through this procedure. Similarly to the phenomenon of superfilamentation in air \cite{Point2014}, plasma channels tend to merge and become undistinguishable around the focus, with an average plasma density exceeding that at the entrance of the focal region. This is in line with recent observations \cite{Potemkin2015} of filamentation in water.

\begin{figure}[h]
\includegraphics[width=\columnwidth]{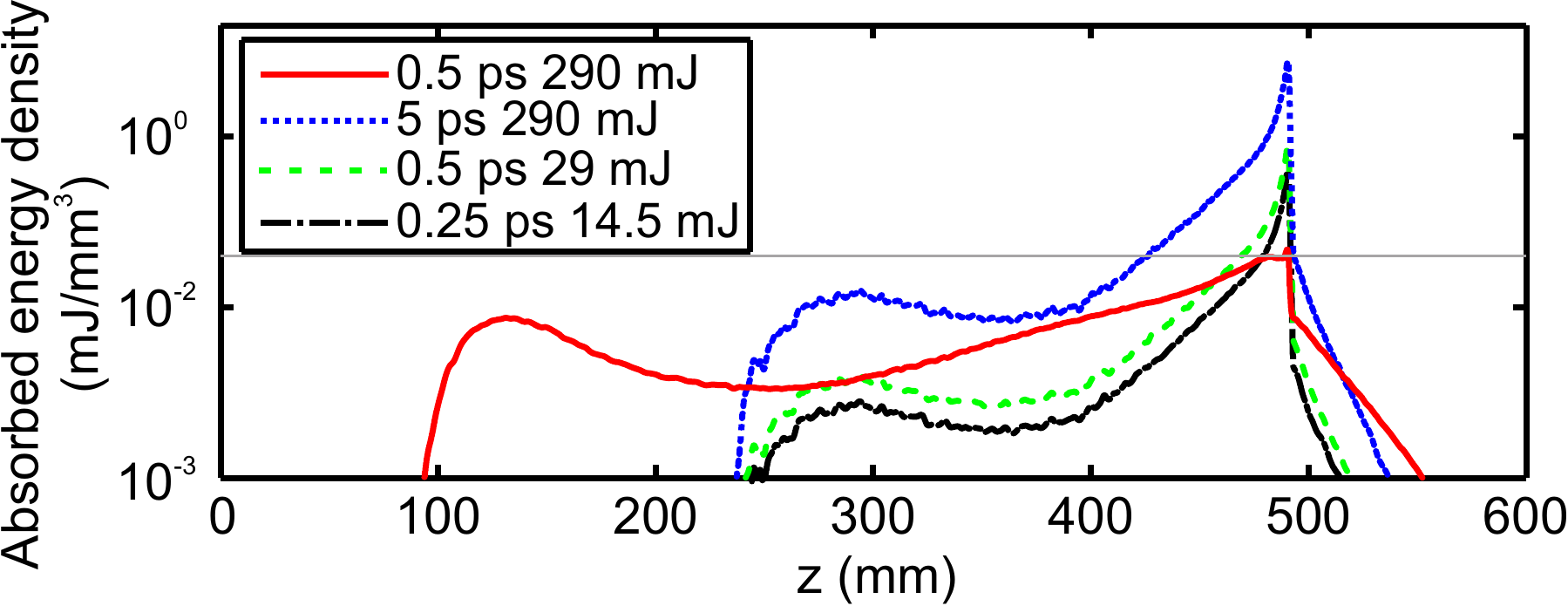}
\caption{\label{fig:AbsorbedEnergyDensity} Absorbed energy density dependence on propagation distance. Different initial beam energies and pulse duration cases are represented with different colors and are the same as in Fig \ref{fig:NumLines1}. The thin gray horizontal line represents an approximate threshold value (40 μJ/mm$^3$) for pressure wave to be registered.}
\end{figure}

The quantitative comparison between the densities of deposited energy $U_m(z)$ for different cases is depicted in figure \ref{fig:AbsorbedEnergyDensity}. It is shown that the density of deposited energy is much larger for 5 ps pulses than in any other case. We note that the sound wave can be recorded by the hydrophone only when the amplitude overcomes the noise level set by the dynamic range of the hydrophone. Therefore we can speculate that the deposited energy density must reach a threshold value to be able to generate an acoustic signal above the detection threshold. If for instance the threshold value is 40 μJ/mm$^3$ - short pulses (0.5 ps) will generate sound waves only at the nonlinear focus where the red curve overcomes the threshold. Therefore for short pulses acoustic waves will be registered coming from a point source, which is similar to the observations. Long pulses (5 ps) overcome the 40 μJ/mm$^3$ threshold value for a large range in the focal region and are expected to generate sound waves above the detection threshold from this extended region.

\begin{figure}[h]
\includegraphics[width=\columnwidth]{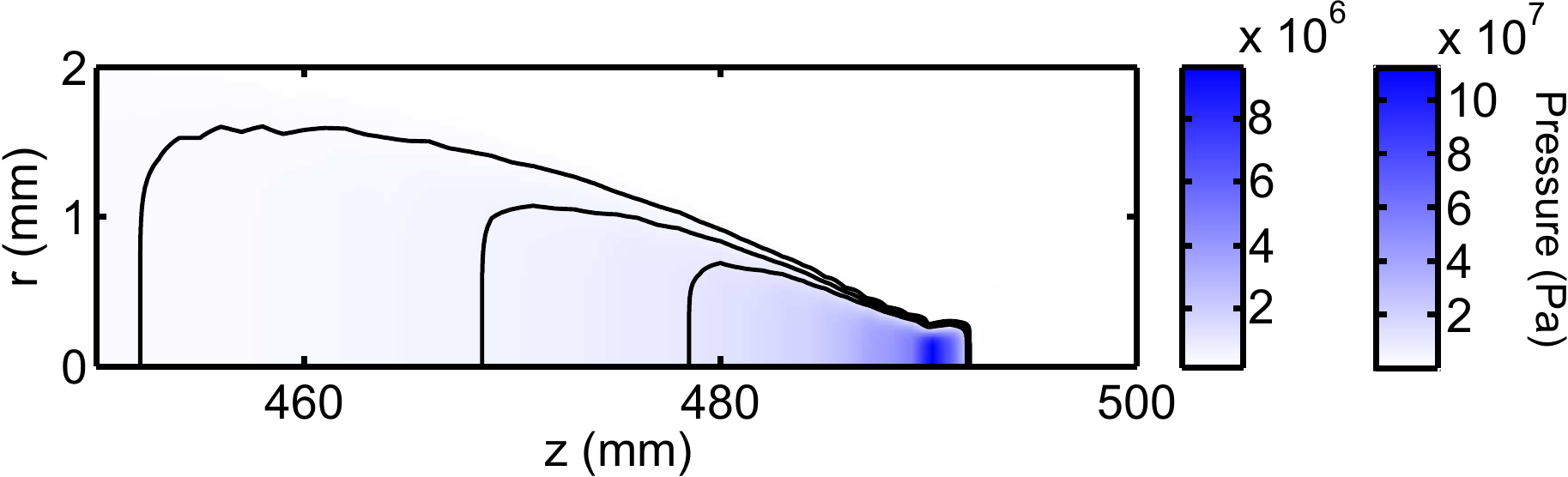}
\caption{\label{fig:InitialPressure} Initial pressure distributions for the peak heatings of $\Delta T$ = 6 K (leftmost colorbar) and $\Delta T$ = 70 K (rightmost colorbar). The contours are at  1/10, 1/20, and 1/50 of the difference between the maximum  pressure and the background level 1.023$\times$10$^5$ Pa, corresponding to a depth of the order of 10 cm in water of background density $\rho_0$= 998.2 kg/m$^3$. Same criterion for contours is applied to Figs. \ref{fig:PressureNumerics} and \ref{fig:AcousticSignal}.}
\end{figure}

From the simulated density of laser energy deposition, we assume cylindrical symmetry to calculate an averaged radial distribution for the elevation of water temperature after the pulse. The equation of state for water links this temperature to the pressure profile that we will use to simulate the generation of an acoustic wave. To account for the steep edges of the averaged profile for the energy deposited in the focal volume, we assumed the latter to be described by a super-Gaussian profile of order $s$ = 4 in Eqs (\ref{eq:Moment},\ref{eq:RUm}), leading to an amplitude of the initial pressure profile depicted in figure \ref{fig:InitialPressure}. Its longitudinal profile follows that of the deposited energy density depicted in figure \ref{fig:AbsorbedEnergyDensity} by red curve (5 ps, 290 mJ). We note that the procedure of radially averaging the deposited energy leads to a maximum elevation of temperature of 6 K above the background. We also performed calculations for a higher elevation of temperature of 70 K, corresponding to a peak pressure of 100 MPa (see rightmost colorbar in Fig. \ref{fig:InitialPressure}), to check whether nonlinearity significantly modifies the acoustic wave.
%%%%%%%%%%%%%%
%    FIG. 9 WAS HERE    %
%%%%%%%%%%%%%%
\section{\label{sec:Acoustic}Numerical simulations of acoustic wave generation and propagation}
\subsection{\label{sec:Linear}Linear acoustics}
In order to interpret the recorded profile of the acoustic signal, we performed simulations of the propagation of acoustic waves by using a simplified model, using the linearized continuity equation and equation of motion,
\begin{eqnarray}
\frac{\partial\rho}{\partial t}+\rho_0\nabla \textbf{v}=0,\nonumber\\
\frac{\partial\textbf{v}}{\partial t}+\frac{1}{\rho_0}\nabla p=0,
\label{eq:Linear}
\end{eqnarray}
together with the equation of state $p=c_s^2\rho$. The model describes linear propagation of acoustic waves. We assumed cylindrical symmetry. As explained above, the initial condition was taken in the form of a simplified pressure profile, shaped as the focal volume where energy was deposited (see Fig. \ref{fig:InitialPressure}). The underlying assumption is that the time required for electron recombination and energy transfer to matter is much shorter than the typical time for initiating the thermo-elastic expansion of water.

For a comparison with experimental results, the pressure wave amplitude was calculated at a fixed distance of 28 mm from the source. This distance ensures that the signal propagated far enough from the source to be considered as a far-field measurement, without a need of expanding further the radial coordinate axis. Results shown below in figure \ref{fig:AcousticSignal} (a) clearly exhibit the same structure as the experimental results plotted in figure \ref{fig:Exp2} (a).  This indicates that the two acoustic branches measured in Fig. \ref{fig:Exp2} (a) originate from a geometric effect associated with the shape of the plasma volume. The geometry of the source of acoustic waves and their linear propagation are sufficient to explain the main features of the signal.
%%%%%%%%%%%%%%%%%%%%%%
% CARLES STARTED RE-TYPING HERE %
%%%%%%%%%%%%%%%%%%%%%%
\subsection{\label{sec:nonlinear}Nonlinear hydrodynamics and acoustic wave generation}

We have investigated numerically the initial expansion of the focal volume after laser energy deposition in order to check if cavitation and shock wave formation significantly affect the diagnostics obtained in section \ref{sec:Linear} by means of a linear acoustics model. Our nonlinear model is based on the compressible Euler equations describing the time evolution of mass density, $\rho$, bulk velocity, $\textbf{v}$, and total energy density, $e$:
\begin{eqnarray}
&&\frac{\partial\rho}{\partial t}+\nabla (\rho\textbf{v})=0\nonumber\\
&&\frac{\partial(\rho\textbf{v})}{\partial t}+\nabla(p\textbf{v}\textbf{v}+p\bf I)=0\nonumber\\
&&\frac{\partial e}{\partial t}+\nabla([e+p]\textbf{v}-\lambda\nabla T)=0,
\label{eq:Euler}
\end{eqnarray}
where $e=\rho\epsilon+\frac{1}{2}\rho |\bf v|^2$ and $\bf I$ is the identity matrix. The system of equations above is closed with the additional expressions for the specific internal energy, $\epsilon(\rho,T)$, and pressure, $p(\rho,T)$, given by the Mie Gr\"{u}neisen equation of state (see Ref. \cite{AkhatovPF2001} for details). Here $\lambda$=0.58  J /[K m s] is the heat flux coefficient and $T$ is the temperature. For waves of small amplitude, Eqns. (\ref{eq:Euler}) reduce to the linear set given by Eqns. \ref{eq:Linear} (see Appendix B).

\begin{figure}[h]
\includegraphics[width=\columnwidth]{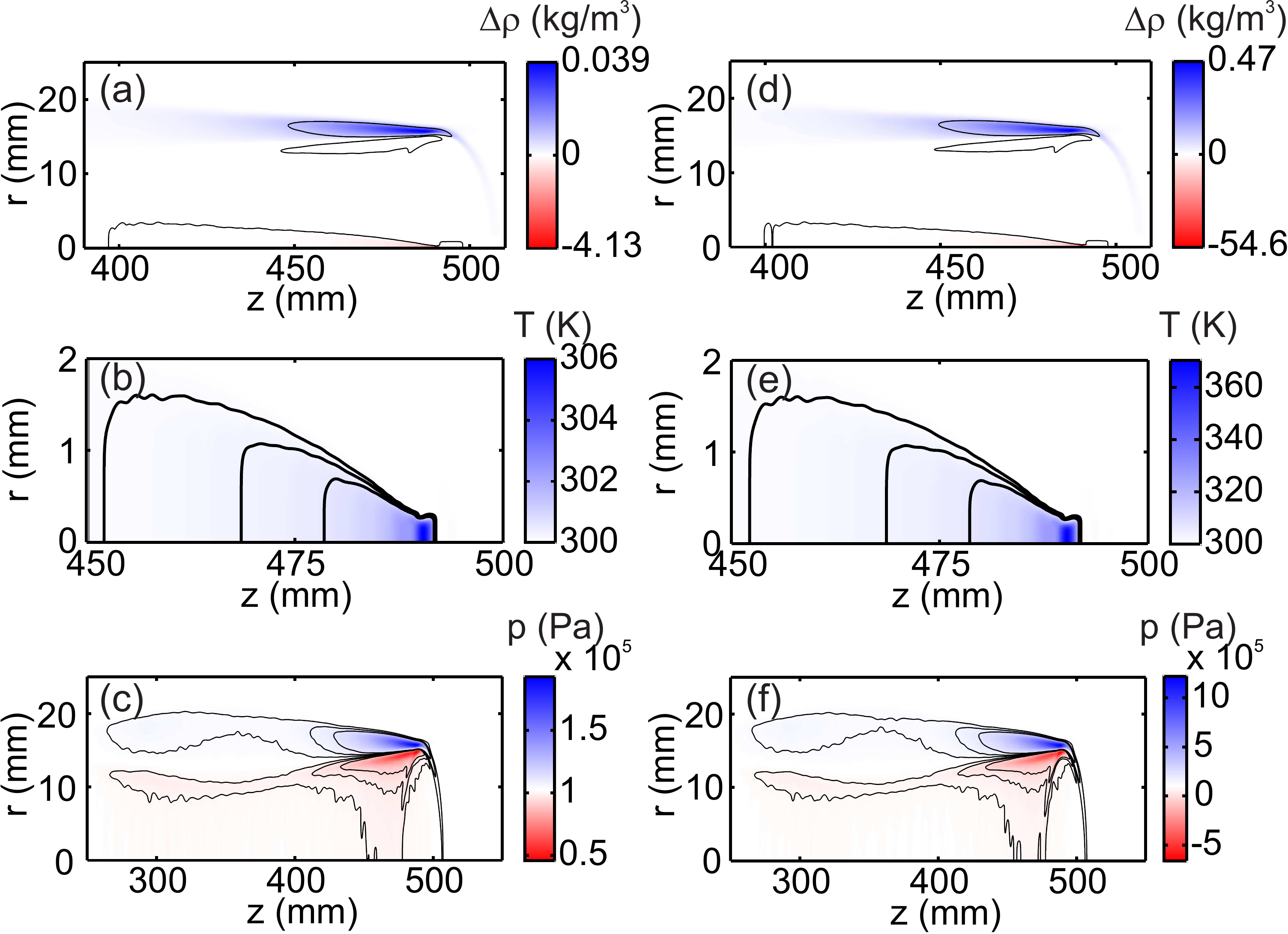}
\caption{\label{fig:PressureNumerics} (a,d) Density $\Delta\rho\equiv\rho-\rho_0$, (b,e) temperature, and (c,f) pressure distributions in water 10 $\mu$s after the pulse heating. (a-c) $\Delta T=6$ K, (d-f) $\Delta T=70$ K. All simulations use cylindrical coordinates with axial revolution symmetry. }
\end{figure}

Equations (\ref{eq:Euler}) are integrated in time, $t$, by means of a hyperbolic solver \cite{ToroPaper}. Figure \ref{fig:PressureNumerics} shows the thermodynamic variables $\rho$, $T$, $p$ 10 $\mu$s after the pulse transit. Our simulations are initialized with $i)$ the pressure distribution $p_0(r,z)$ shown in Fig. \ref{fig:InitialPressure}, $ii)$ the equilibrium density $\rho_0$= 998.2 kg/m$^3$ everywhere, since the medium barely moves during the pulse transit times $\ll$ns, $iii)$ $\bf v(t=0)=\bf 0$, consistent with $ii)$, and $iv)$ with $e(\rho_0,p_0)$ given by the equation of state.

Figure \ref{fig:PressureNumerics} presents results obtained by assuming initial peak temperatures $\Delta T\equiv \max\{T(r,z,t=0)-T_0\}=$6 K (left) and 70 K (right) above the room temperature $T_0$ = 300 K. %We observe that the temperature profile remains practically constant in time, as expected from the large characteristic times for heat diffusion in water (hundreds of $\mu$s). In contrast, the large pressure in the focal volume develops a dip in density and an acoustic wave that detached from the focal region that propagates mainly along the radial direction.
The initial stages are characterized by a fast evolution of density and pressure. This is due to the ultrafast energy deposition from the laser source to the medium, which occurs at constant density rather than at a constant pressure (i.e, in mechanical equilibrium). Heating of the focal volume occurs while plasma recombines, much faster than the hydrodynamic time-scales for diffusion or fluid motion and therefore the system is driven out of the equilibrium. Immediately after the heating of the focal volume, the temperature remains almost constant in time due to the very low heat conductivity ($T$ relaxes over the scale of ms in water). Under these conditions, the density of water rapidly drops below the background level as pressure decreases to restore the mechanical equilibrium (flat $p$) around the laser focus. This transient process is indeed the responsible for the emission of an acoustic wave. In the far-field, only the amplitude of the acoustic wave differs but the wave profile is similar for both heating levels.

\begin{figure}[h]
\includegraphics[width=\columnwidth]{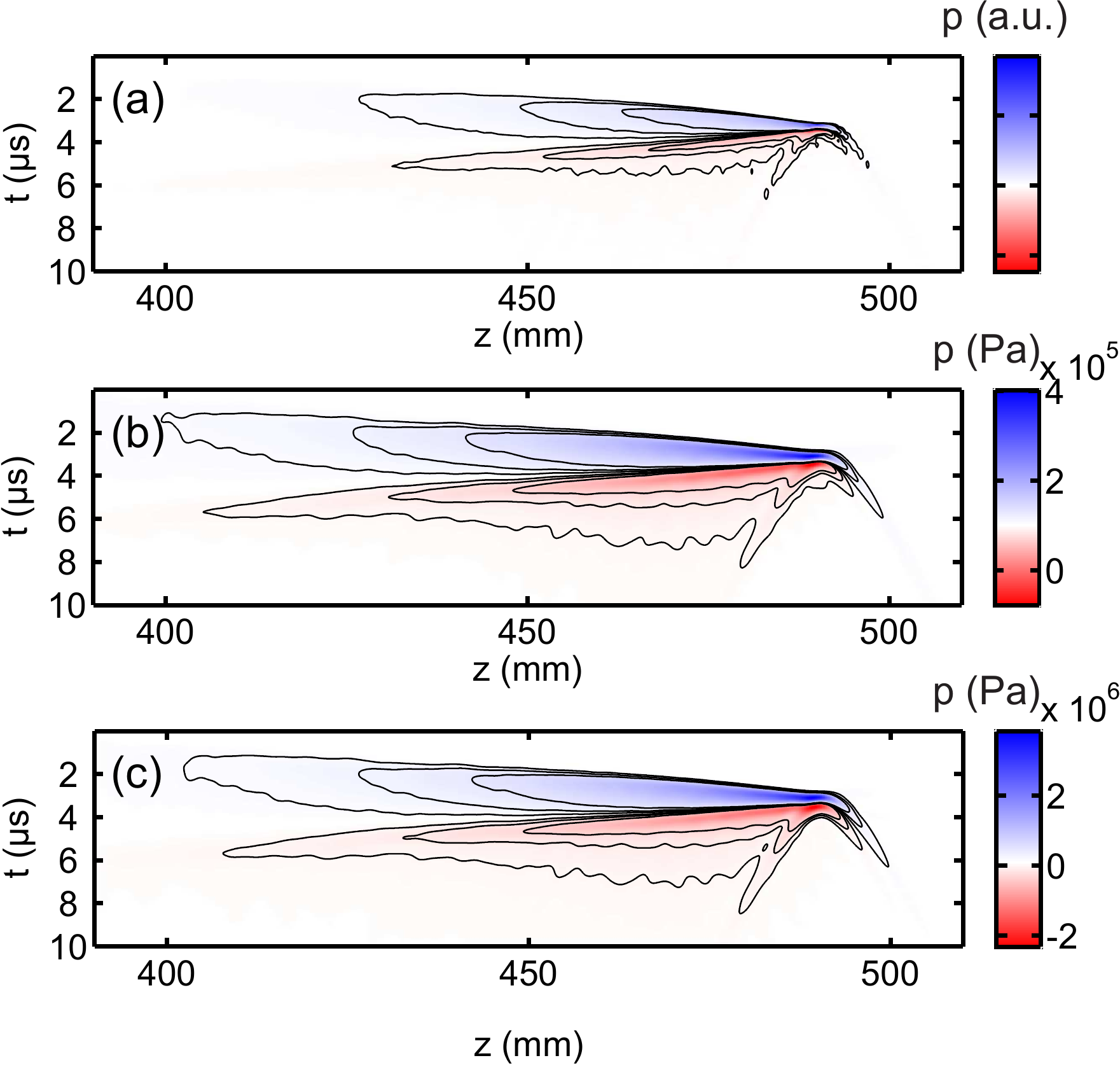}
\caption{\label{fig:AcousticSignal} Simulated pressure vs time signal recorded at an offset of 5 mm from the pulse propagation axis, $z$. Simulations show (a) linear (Eqns. \ref{eq:Linear}) and (b,c) nonlinear (Eqns. \ref{eq:Euler}) calculations, the latter for (b) $\Delta T= 6$ K and (c) $\Delta T= 70$ K.}
\end{figure}

\begin{figure}[h]
\includegraphics[width=\columnwidth]{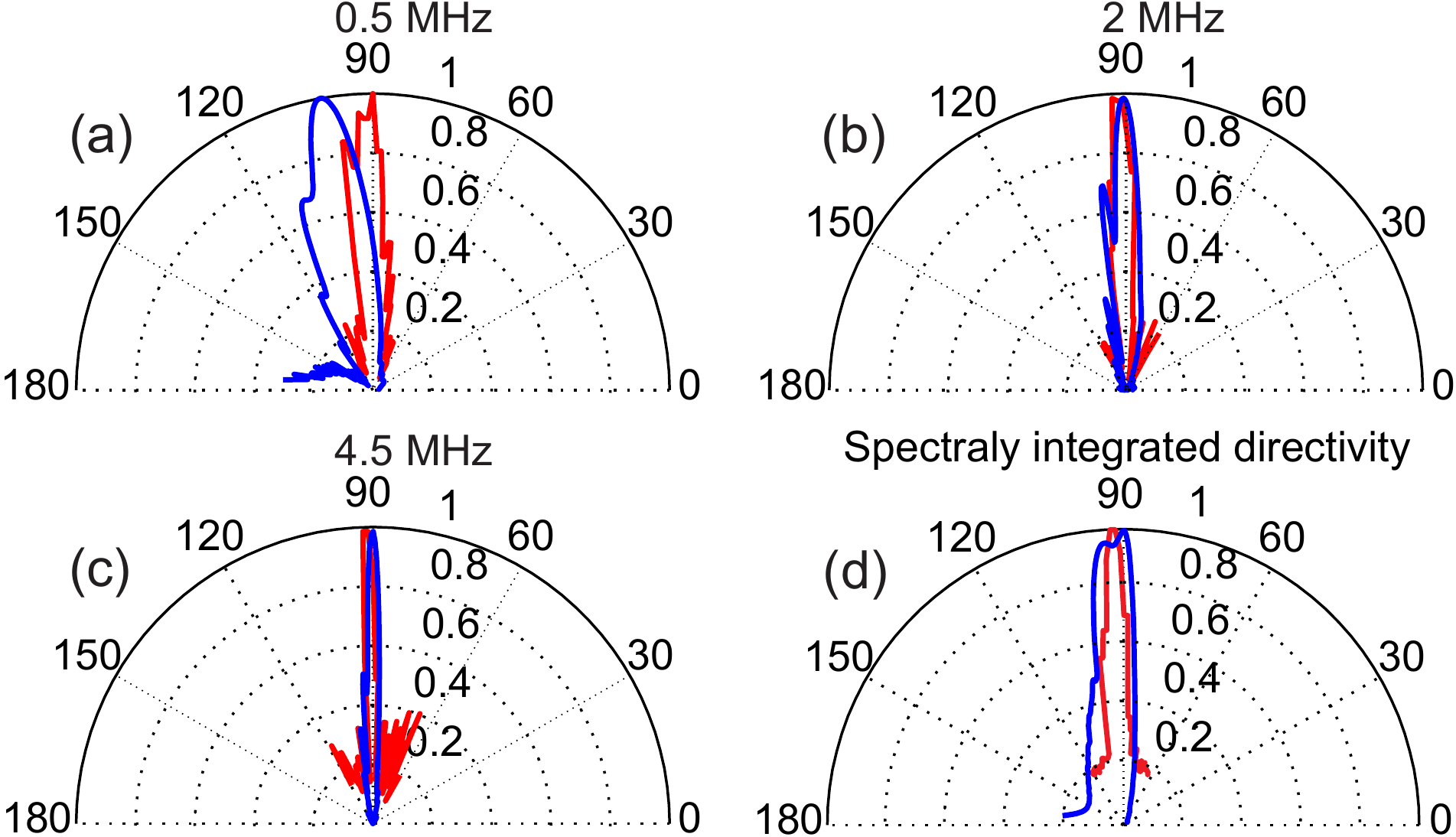}
\caption{\label{fig:Directivity} Sound wave directivity measured from experiments (red curves) and numerical simulations with $\Delta=70$ K (blue curves) for (a) 0.5 MHz, (b) 2 MHz, and (c) 4.5 MHz. The maximum pressure recorded numerically at this distance is of $2$ MPa above the background $\sim0.1$ MPa and the maximum amplitudes for the selected frequencies correspond to (a) 5.3$\times$10$^5$ Pa/MHz, (b) 2.88$\times$10$^5$  Pa/MHz, and (c) 0.98$\times$10$^5$  Pa/MHz. (d) Shows spectrally integrated angular distribution. Angles are measured from the z axis: 0$^{\circ}$ (180$^{\circ}$) correspond to the forward (backward) laser propagation direction.}
\end{figure}

Figure \ref{fig:AcousticSignal} shows a comparison of the temporal profile for the acoustic signals that would be captured by a hydrophone placed 5 mm off axis, for linear and nonlinear simulations initiated with different over-pressures. The lower is the initial overpressure, the closer the agreement is expected to be between Eqns. (\ref{eq:Euler}) and (\ref{eq:Linear}). However, results are close for heating levels up to 70 K above the background temperature. We observe that the profiles of the acoustic waves simulated with the compressible Euler equations (Figs. \ref{fig:AcousticSignal}b,c) are very close to that obtained by linear acoustics (Fig. \ref{fig:AcousticSignal}a), and all are in good qualitative agreement with the measured profile (Fig. \ref{fig:Exp2}c). These results confirm that the geometry of the source of acoustic waves and their linear propagation are sufficient to explain the main features of the signal. This cannot be granted in general: while perturbations of water in equilibrium are small, the isothermal compressibility is very high (for example, at 300 K and 1 atm., the isothermal compressibility $\beta_T\equiv 1/\rho(\partial\rho/\partial p)_T\approx10^{-10}Pa^{-1}$: an increase of pressure of 1 MPa, $\sim10$ atm, will change the water density in 1 part amongst 10.000).
%(\textcolor[rgb]{1.00,0.00,0.00}{isothermals on the  ($p$,$\rho$) plane are very steep and thus big variations of pressure yield small variations of density}\textbf{Needs evaluation of typical numbers of steepness(gradient)?});
Additionally, even a weak localized heating can easily induce cavitation and phase changes, reducing compressibility dramatically. Thus, simulations carried out with the compressible Euler equations allowed us to check that there is little difference in the acoustic wave propagation once it is detached from the focal region. The two branches in the far-field acoustic signal in Fig. \ref{fig:Exp1} originate essentially from the shape of the plasma volume where laser energy is deposited.
We also performed a numerical directivity study. We recorded pressure vs time on 300 virtual microphones evenly distributed over the 50 mm radius half circumference centered at the position of the maximum initial pressure ($r=0$, $z\approx490$ mm). In figure \ref{fig:Directivity} a comparison with experimental results is shown for the angular dependence of the amplitude for selected frequencies. Most of the sound wave energy is distributed perpendicularly to the beam propagation axis, as measured and shown by Y. Brelet et al. \cite{Brelet2015}. We note that the directivity is sharply peaked for higher frequencies. This is associated with the fact that the larger volume of the conically shaped initial pressure profile generates lower frequencies and therefore, the directivity is looser, while the high frequency components are generated at the elongated but thin peak. A thin initial pressure profile generates steeper pressure fronts with broader spectrum. Slight discrepancy for the directivity at low frequencies is due to the closer position of microphones in the numerical simulations.
%\begin{figure}[h]
%\includegraphics[width=\columnwidth]{Directivity2.jpg}
%\caption{\label{fig:Directivity2} Pressure wave spectral directivity.   }
%\end{figure}

\section{Conclusion}
We have demonstrated a high directivity acoustic source generated by loosely focused multi milijoule picosecond pulses in water. The acoustic wave predominantly propagates transversally to the laser beam and its origin is attributed to the phenomenon of superfilamentation. The dependence of the acoustic signal upon pulse duration is numerically and experimentally investigated. While fs pulses tend to produce a point acoustic source and energy deposition per unit volume remains relatively low, ps pulses produce an extended source and deposit much more energy per unit volume, yielding very high directivity and higher power. The laser induced hydrodynamics is fully studied numerically by means of the compressible Euler equations and a suitable equations of state for water. Additionally, a simplified linear acoustic model provides efficient calculations of the pressure far fields, linking the calculations with the experimental measurements. The combination of optical and hydrodynamical results are in an overall good agreement with experiments. Our findings are relevant for underwater detection and communications, where the ability to dynamically control the directivity of the acoustic sources is important.

%We have demonstrated acoustic wave generation with multi milijoule ultrashort pulses. The dependence of the acoustic signal upon pulse duration is presented. The conically shaped spatio-temporal acoustic signal generated with picosecond pulse is recorded experimentally, while with fs pulses, the acoustic signal comes from the point source located in the vicinity of the linear beam focus. Numerical simulation results show a dependence of laser energy deposition on laser pulse duration. The deposited energy density is much higher for ps pulses and energy is deposited into a smaller volume. The initial overpressure was calculated by means of equations of state for water. The generation of acoustic wave was calculated numerically by means of compressible Euler equations showing the emergence of a conically shaped spatio-temporal pressure profile. A linear acoustic model led to similar results for the parameters of our experiments. The acoustic wave was calculated to propagate perpendicularly to the beam propagation axis, exhibiting a high directivity, in very good agreement with our measurements.

\begin{acknowledgments}
This project has been supported by the French Direction Generale de l$'$Armement (Grant numbers 066003600470750, 2012.60.0013 and 2015.60.0004). The authors thank the staff from Laboratoire de Mecanique et Acoustique for technical assistance.
\end{acknowledgments}
\appendix
\section{}
The propagation equation used for laser beam filamentation:
\begin{eqnarray}
  \frac{\partial E}{\partial z}&=&\frac{i}{2n_0k_0}\nabla^2_\bot E+ik_0n_2|E|^2E-\frac{\sigma}{2}(1+i\omega_0\tau_c)\rho_eE\nonumber\\
  &&-\frac{\beta_K}{2}|E|^{2K-2}\left(1-\frac{\rho_e}{\rho_{nt}}\right)E,\nonumber\\*
  \frac{\partial\rho_e}{\partial t}&=&\frac{\beta_K}{K\hbar\omega_0}|E|^{2K}\left(1-\frac{\rho_e}{\rho_{nt}}\right)+\frac{\sigma}{U_i}\rho |E|^2.
\end{eqnarray}
Here $n_0$=1.33 is refractive index, $n_2$=19$\times$10$^{-21}$ m$^2$/W - nonlinear refractive index, $\sigma$=4.7$\times$10$^{-22}$ m$^2$ - cross section for inverse Bremsstrahlung, $\tau_c$=3 fs - electron collision time, $\beta_K$=8.3$\times$10$^{-52}$ cm$^7$/W$^4$ - multiphoton absorption (MPA) coefficient, $K$=5 - MPA order, $\rho_{nt}$=6.7$\times$10$^{22}$ cm$^{-3}$ - neutral atom density and $\rho_e(x,y,z,t)$ - plasma density.
\section{}
In the limit of low isothermal compressibility, $\beta_T\equiv\frac{1}{\rho}\left.\frac{\partial\rho}{\partial p}\right\vert_T\ll p^{-1}$ (or $\frac{\partial\rho}{\partial t}\ll\frac{\rho}{p}\frac{\partial p}{\partial t}$) and provided that $c_s^2\sim p/\rho$, eqns. (\ref{eq:Euler}) reduce to the linear set eqns. (\ref{eq:Linear}) where slow variations of density are assumed. In our case, the incompressibility limit states that the Mie Gr\"{u}neisen speed of sound is given by $c_s^2=[ap+b]/\rho$ ($a \sim$ 20 and $b \sim$ 10$^{10}$ Pa for $T \sim$ 300 K) and the Euler equations therefore reduce to
\begin{equation}
\frac{\partial^2p}{\partial t^2}=\frac{c^2}{a}\nabla^2\left(p\left[1+\frac{au^2}{c^2}\right]+b\frac{u^2}{c^2}\right).
\end{equation}
Given the typical values of $p$ $\sim$ 10$^7$ Pa, and $u/c$ $\sim$ 10$^{-2}$, propagation is given by the linearized form $\frac{\partial^2p}{\partial t^2}=c'^2\nabla^2p$ with $c'=\sqrt{\frac{b}{a\rho}}\sim 10^3$, m/s for typical densities of water $\sim$10$^3$ kg/m$^3$.

%%%%%%%%%%%%%%%%%%%%%%
% CARLES STOPPED TYPING HERE %
%%%%%%%%%%%%%%%%%%%%%%

%%%%%%%%%%%%%%%%%%%%%%%%%%%%%%%%%%%%%%%%%%%%%%%%%%%%%%%%%%%%%%%%%%%%%%%%%%%%%%%%%%%%%%%%%%%%%

\nocite{*}

\bibliography{waterPubl}% Produces the bibliography via BibTeX.

\end{document}